# Giant Tunable Faraday Effect in a Semiconductor Magneto-plasma for Broadband Terahertz Polarization Optics

**Takashi Arikawa,[1] Xiangfeng Wang,[1] Alexey A. Belyanin,[2] and Junichiro Kono[1],***

[1]*Departments of Electrical & Computer Engineering and Physics & Astronomy, Rice University, 6100 Main St., Houston, Texas 77005, USA*
[2]*Department of Physics, Texas A&M University, 301 Tarrow Street, College Station, Texas 77843, USA* *[*]kono@rice.edu*

**Abstract:** We report on a giant Faraday effect in an electron plasma in *n*-InSb probed via polarization-resolved terahertz (THz) time-domain spectroscopy. Polarization rotation angles and ellipticities reach as large as $\pi/2$ and 1, respectively, over a wide frequency range (0.3-2.5 THz) at magnetic fields of a few Tesla. The experimental results together with theoretical simulations show its promising ability to construct broadband and tunable THz polarization optics, such as a circular polarizer, half-wave plate, and polarization modulators.

## 1. Introduction

Recent advancements in terahertz (THz) technology [1] – for generation [2,3], manipulation [4,5], and detection [6] of electromagnetic radiation in the 0.1-10 THz frequency range – have enabled an impressive array of basic studies [7] and applications [8] in spectroscopy, imaging, security screening, and high-speed communications. However, compared to the near-infrared and visible range, some of the critical components are still not as sophisticated, and, in some cases, simply absent. One poorly developed area is active THz polarization optics, which would allow fast polarization modulations for e.g., real-time ellipsometric imaging [9,10] and vibrational circular dichroism of biological molecules [11,12].

The challenge is that conventional materials and principles traditionally employed in the visible and infrared region do not always provide required functionalities in the THz region. Hence, recent studies have proposed novel schemes based on artificial dielectrics and metamaterials for realizing THz quarter-wave plates [13,14]. They exhibit excellent performances at a single frequency, suitable for narrowband or continuous-wave applications. However, since the operation frequency depends on the physical dimensions of the material structure, these schemes cannot be extended to broadband operation. Considering the wide use of broadband THz pulses in time-domain spectroscopy, an alternative principle to realize broadband polarization optics is strongly desired. An achromatic quarter-wave plate has been demonstrated based on multilayer quartz [15]. There are a few previous reports on the generation of THz pulses with arbitrary elliptical polarization [16-18], which allow polarization modulations by mechanically changing the time delay between two optical pulses. A broadband polarization modulator between right- and left-circularly polarized modes has also been demonstrated by utilizing a four-contact photoconductive antenna and linear-to-circular-polarization converter [19].

Here, we demonstrate a novel broadband and tunable THz polarization optics based on a giant Faraday effect in electron-doped InSb crystals. The effect derives from the material's different complex indices of refraction for right- and left-circularly polarized light (cyclotron resonance active, CRA, and cyclotron resonance inactive, CRI, respectively [20]) in an applied magnetic field and is demonstrated to work as a broadband circular polarizer and half-wave plate. Our method is applicable to any types of THz sources, while the previous methods are THz source-specific. Furthermore, the operation principle demonstrated here will eventually allow us to modulate the polarization states of THz radiation electronically, which opens up the possibility of developing a fast polarization modulator, similar to photoelastic modulators commonly used in the visible and infrared spectral region.

## 2. Experiment

We used THz time-domain magneto-spectroscopy in a transmission geometry (Fig. 1a) [21]. Coherent THz pulses were generated from a nitrogen gas plasma [22] created by focusing both the fundamental and the second harmonic wave of the output of a chirped-pulse amplifier (CPA-2001, Clark-MXR, Inc) with a center wavelength of 775 nm, a pulse width of ~150 fs, a pulse energy of ~600 µJ, and a repetition rate of 1 kHz. A BBO crystal was placed between the lens ($f$ = 10 cm) and the focus position. The distance from the lens to the BBO crystal was 8 cm. The emitted THz pulses were elliptically polarized. The first wire grid polarizer (WGP) was placed before focusing the THz wave to the sample to make the incident THz wave linearly polarized along the $x$ axis. The extinction ratio of the WGP was $10^{-3}$ in power. The transmitted THz electric field through the sample was directly measured by an electro-optic sampling method using a (110) ZnTe crystal (1 mm thick) [23]. To measure only the $y$-component of the transmitted THz wave, the second WGP was placed in cross-Nicole geometry before the ZnTe crystal, whose [001] axis was perpendicular to the polarization of the probe beam ($y$-polarized). In this geometry of the ZnTe crystal and probe beam, the $x$-component of the THz wave did not induce any electro-optic signal [23], which assured an accurate determination of a $y$-component in the presence of the leakage transmission of the $x$-component through the second WGP. The $x$-component was measured by setting the second WGP parallel to the first WGP and the [001] direction of the ZnTe crystal parallel to the probe beam. In this case, the $y$-component of the THz wave did not contribute to the electro-optic signal. The two geometries of electro-optic sampling described above had the same sensitivity to the $y$- and $x$-components of the THz electric field [23], enabling quantitative comparison between the two. The frequency bandwidth was from 0.3 to 2.5 THz. All the experiments were done under nitrogen purging to avoid water vapor absorption in ambient air. The sample was cooled down to 184 K in an optical cryostat with a superconducting magnet (SM4000-10, Oxford instruments). The Faraday geometry was used, where both the propagation direction of the THz wave and the magnetic field direction were perpendicular to the sample surface.

The sample was a Te-doped $n$-InSb crystal (20 mm x 30 mm x 0.63 mm) with an electron density of $6.1 \times 10^{14}$ cm$^{-3}$. At 184 K, the thermally excited electron density was estimated to be ~ $1.5 \times 10^{14}$ cm$^{-3}$.

## 3. Results

Narrow-gap compound semiconductors such as InSb, InAs, and HgTe are known to show giant free-carrier Faraday rotation in the far-infrared and microwave range [24,25]. The magneto-optical properties of InSb have been studied for decades [20,24,26,27], but broadband Faraday rotation spectroscopy measurements have not been reported. Figure 1b shows the $x$- and $y$-components of the transmitted THz pulses [$E_x(t)$ and $E_y(t)$] through the sample at 184 K at several magnetic field ($B$) values. We observed $y$-components with similar amplitudes to the x-components in a field as low as 0.1 T, demonstrating a giant Faraday effect. To visualize the polarization state, we made parametric plots as shown in Fig. 2. Each point represents the tip of the electric field vector at each time. At 0 T, the transmitted THz field remains x-polarized, while at fields up to 1 T, it is close to circular polarization [28]. At 1.5 T and 2 T, the trajectory is distorted and becomes elliptically polarized. Finally, above 2.5 T, it becomes linearly polarized again, but the oscillation direction of the electric field is significantly rotated.

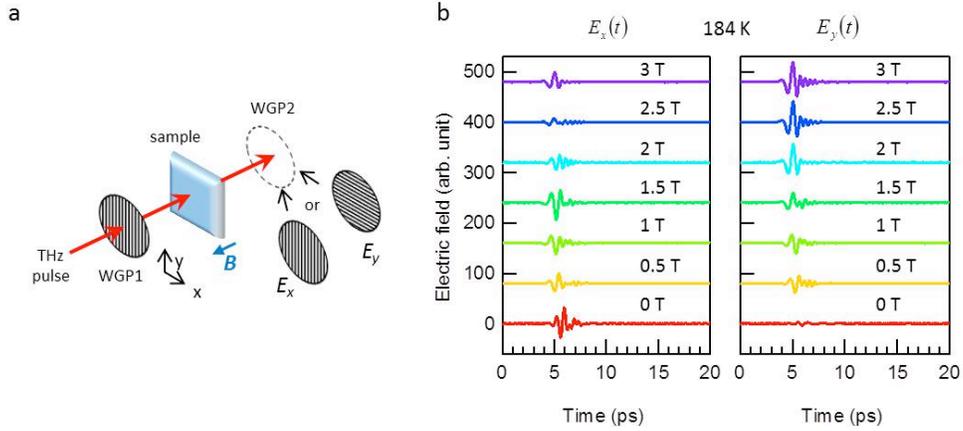

Fig. 1. Faraday effect in $n$-InSb. a) Schematic of the experimental setup. WGP, wire grid polarizer. b) The $x$- (left) and $y$-components (right) of the transmitted THz pulses through the sample at selected magnetic fields $B$. Traces are vertically offset for clarity. The temperature of the sample was 184 K.

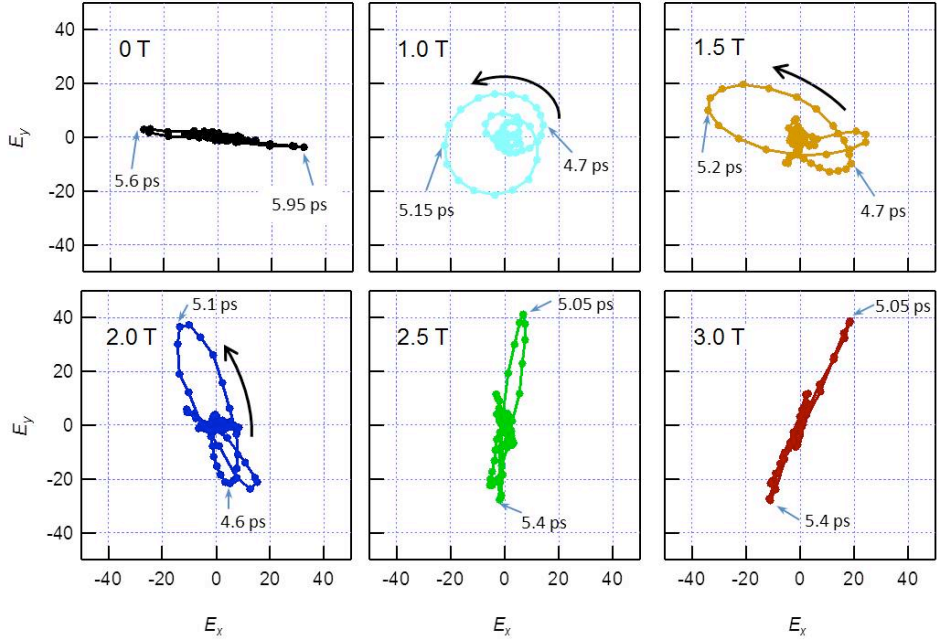

Fig. 2 Polarization state of the transmitted THz pulses. Parametric plots of $E_x$ and $E_y$. The time interval between the dots is 50 fs. The transmitted THz pulses are almost circularly polarized (counterclockwise) in the lower magnetic field regime ($B \leqq 1$ T). The transmitted THz pulses are elliptically polarized (counterclockwise) in the intermediate regime (1.5 T and 2.0 T). The transmitted THz pulses are becoming linearly polarized in the higher magnetic field regime (2.5 T and 3.0 T).

To quantitatively describe the observed Faraday effect, we determined Faraday ellipticities and rotation angles in the frequency domain (dotted lines in Fig. 3) using the data in Fig. 1. Faraday rotation angle $\theta$ and ellipticity $\eta$ are defined as follows,

$$\theta(\omega) = \frac{\arg\{\tilde{E}_{CRA}(\omega)\} - \arg\{\tilde{E}_{CRI}(\omega)\}}{2} \quad (1)$$

$$\eta(\omega) = \frac{|\tilde{E}_{CRI}(\omega)| - |\tilde{E}_{CRA}(\omega)|}{|\tilde{E}_{CRI}(\omega)| + |\tilde{E}_{CRA}(\omega)|}. \quad (2)$$

Here, $\tilde{E}_{CRA}(\omega)$ and $\tilde{E}_{CRI}(\omega)$ are the complex electric-field amplitudes of the CRA and CRI modes, respectively. These are obtained from the following equations,

$$\tilde{E}_{CRA}(\omega) = \frac{\tilde{E}_x(\omega) + i\tilde{E}_y(\omega)}{\sqrt{2}} \tag{3}$$

$$\tilde{E}_{CRI}(\omega) = \frac{\tilde{E}_x(\omega) - i\tilde{E}_y(\omega)}{\sqrt{2}}, \tag{4}$$

where $\tilde{E}_x(\omega)$ and $\tilde{E}_y(\omega)$ are the complex Fourier transforms of $E_x(t)$ and $E_y(t)$ shown in Fig. 1, respectively.

As shown in Fig. 3, below 2.5 T, Faraday ellipticities reach 1 around the cyclotron resonance (CR) frequencies $\omega_c = 2\pi f_c = eB/m_e c$ where $e$ is the electronic charge, $m_e \approx 0.018 m_0$ is the effective mass of electrons, and $m_0$ is the free electron mass. This means that the transmitted light is a circularly-polarized CRI mode [see Eq. (2)] because the CRA mode is completely absorbed by free carriers. Accordingly, in the frequency region where the ellipticity is unity, the rotation angle cannot be defined. At 3 T, the CR frequency is much higher than the experimental frequency window, resulting in the almost zero ellipticity and dispersionless rotation angle.

## 4. Cold magneto-plasma model simulation

We used the cold magneto-plasma model to describe the observed Faraday effect. Assuming that the incident light is linearly polarized along the $x$-axis, $\theta$ and $\eta$ can be calculated by the following equations,

$$\theta(\omega) = \frac{\arg\{\tilde{T}_{CRA}(\omega)\} - \arg\{\tilde{T}_{CRI}(\omega)\}}{2} \tag{5}$$

$$\eta(\omega) = \frac{|\tilde{T}_{CRI}(\omega)| - |\tilde{T}_{CRA}(\omega)|}{|\tilde{T}_{CRI}(\omega)| + |\tilde{T}_{CRA}(\omega)|}, \tag{6}$$

where $\tilde{T}_{CRA}(\omega)$ and $\tilde{T}_{CRI}(\omega)$ are the complex transmission coefficients of $n$-InSb for the CRA and CRI modes, respectively. These are calculated by the following formulae,

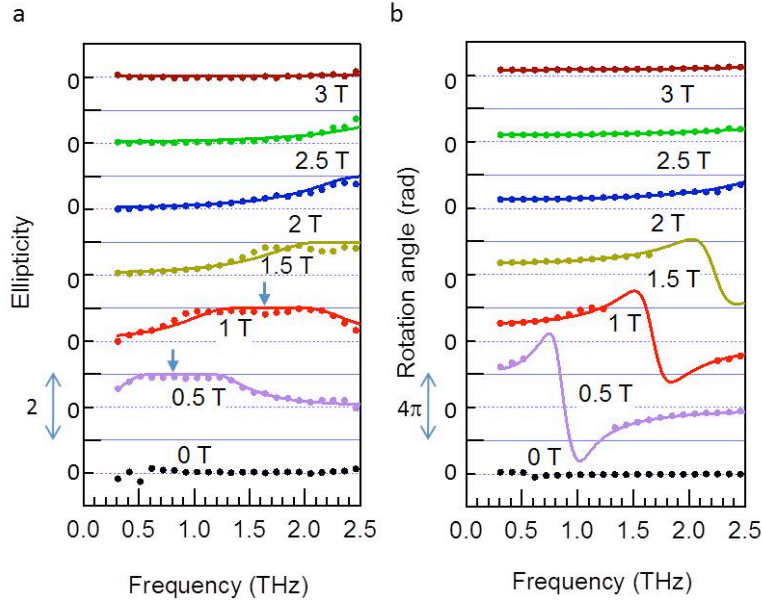

Fig. 3 Faraday ellipticity and rotation angle. The dots show experimental data and the solid lines show fitting results with the cold magneto-plasma model. a) The magnetic field dependence of the ellipticity. The down arrows show the CR frequency at each magnetic field. b) The magnetic field dependence of the rotation angle. The data points are not shown in the frequency region where the corresponding ellipticity is one because the rotation angle cannot be defined.

$$\tilde{T}_{CRA(CRI)} = \tilde{t}^{as}_{CRA(CRI)} \tilde{t}^{sa}_{CRA(CRI)} \exp\left\{i\left(\tilde{n}_{CRA(CRI)} - 1\right)\omega d/c\right\} \quad (7)$$

where $\tilde{n}_{CRA(CRI)}$, $d$ and $c$ are the complex refractive index of $n$-InSb for the CRA (CRI) mode, thickness of the sample, and the velocity of light in vacuum, respectively. $\tilde{t}^{as}_{CRA(CRI)}$ and $\tilde{t}^{sa}_{CRA(CRI)}$ are the Fresnel transmission coefficients from air to the sample and from the sample to air, respectively. The refractive indexes for the CRA and CRI modes can be obtained through the following formulae,

$$\tilde{n}^2_{CRA} = \tilde{\varepsilon}_{xx} - i\tilde{\varepsilon}_{xy} = \varepsilon_b - \frac{\omega_e^2}{\omega(\omega + iv_e - \omega_{Ce})} - \frac{\omega_h^2}{\omega(\omega + iv_h + \omega_{Ch})} + \tilde{\varepsilon}_{ph}, \quad (8)$$

$$\tilde{n}^2_{CRI} = \tilde{\varepsilon}_{xx} + i\tilde{\varepsilon}_{xy} = \varepsilon_b - \frac{\omega_e^2}{\omega(\omega + iv_e + \omega_{Ce})} - \frac{\omega_h^2}{\omega(\omega + iv_h - \omega_{Ch})} + \tilde{\varepsilon}_{ph}, \quad (9)$$

where we used the complex dielectric tensor elements for a cold magneto-plasma [26,29,30]. Kinetic effects were insignificant at our temperatures and frequencies. The phonon contribution is added in the harmonic oscillator approximation [31]: $\tilde{\varepsilon}_{ph} = (\omega_t^2 - \omega_l^2)\varepsilon_b / (\omega_t^2 - \omega^2 - i\nu_{ph}\omega)$ where the parameters $\omega_t, \omega_l$ and $\nu_{ph}$ (divided by $2\pi$) are 5.90, 5.54, and 1 THz, respectively. This contribution is small but non-negligible. Other parameters are as follows: $\varepsilon_b = 16$ is the background permittivity, $\omega_{Ce,h} = |e|B/(m_{e,h}c)$ the cyclotron frequency, and $\omega^2_{e,h} = 4\pi e^2 N_{e,h}/m_{e,h}$ the plasma frequencies for electrons and holes, where the electron and hole densities included both the doping density of electrons and thermally excited densities of electrons and holes. We included both the temperature dependence of the effective band gap, which enters the expressions for thermally generated carrier densities, and the conduction band non-parabolicity. We found that electron masses that provide the best fit to the data at 184 K grow with magnetic field slightly faster than that predicted by the Kane model in [26]. Therefore, we treated both the electron mass and the electron scattering rate ν as fitting parameters when calculating the dielectric tensor for each magnetic field. The electron mass varied from 0.017 to 0.020 $m_0$ with increasing magnetic field from 0.5 to 3 T. The electron scattering rate $\nu_e$ varied between $6 \times 10^{11}$ s$^{-1}$ and $1.7 \times 10^{12}$ s$^{-1}$. The hole scattering rate $\nu_h$ was assumed to be constant at $6 \times 10^{13}$ s$^{-1}$, since the contribution due to holes was small.

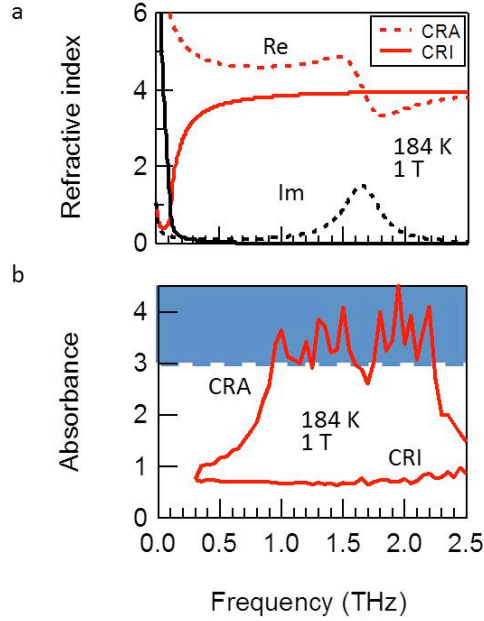

Fig. 4 Optical constants of *n*-InSb for CRA and CRI modes at 1 T. a) The complex refractive indexes for CRA and CRI modes determined by cold magneto-plasma model fitting. The dispersion in the CRA mode around 1.65 THz is due to CR. The dispersions both in the CRA and CRI modes in the low frequency region around 0.1 THz are due to the magneto-plasma resonance. b) The absorbance for CRA(CRI) modes, $A_{CRA(CRI)} = -\log_{10}[T_{CRA(CRI)}]$, where $T_{CRA(CRI)}$ is the transmittance defined as $T = |\tilde{E}_{CRA(CRI)}/\tilde{E}_{ref}|^2$. $\tilde{E}_{ref}$ is the complex electric-field amplitudes in the frequency-domain obtained without a sample. The broken horizontal line represents the maximum absorbance (~3) measurable with our system. The oscillations in the absorbance from 0.9 to 2.3 THz is due to noise and meaningless.

The excellent agreement between the data and theory (solid lines in Fig. 3) suggests that the observed Faraday effect is due to free carriers. One can determine the electron effective mass and scattering time, and calculate the complex refractive indexes of $n$-InSb for CRA and CRI modes as shown in Fig. 4a for $B = 1$ T. One can see some dispersion due to CR at ~1.65 THz in the index for the CRA mode, while there is no dispersion in the CRI mode. The dispersion in the low frequency region around 0.1 THz is due to the magneto-plasma resonance.

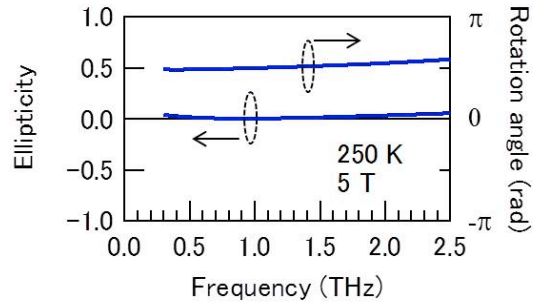

Fig. 5 Theoretical plot of the ellipticity and rotation angle at 250 K. The external magnetic field is 5 T and the thickness of the crystal is 0.3 mm. The rotation angle (ellipticity) is close to p/2 (zero) from 0.3 to 2.5 THz.

## 5. Discussion

This giant Faraday effect provides us with a simple way to realize broadband THz polarization optics. This scheme utilizes the magnetic-field-induced anisotropy in the optical constants between CRA and CRI modes. This is in stark contrast to usual wave plates, where the difference in the optical constants between two orthogonal linearly-polarized modes along the fast- and slow-axes is utilized. Because the optical anisotropy between the CRA and CRI modes can be easily tuned by the external magnetic field, we can use $n$-InSb as both "quarter-" and "half-wave plates." Let us first discuss the characteristics of $n$-InSb as a broadband "quarter-wave plate," or more properly, circular polarizer. As shown above, $n$-InSb works as a filter that absorbs the CRA mode completely around the CR frequency; i.e., the transmitted light is purely in the CRI mode. The bandwidth is defined as the frequency region where the field strength of the transmitted CRA mode is below the dynamic range of the experimental system. Figure 4b shows the absorbance of the CRA and CRI modes at 1 T together with the maximum absorbance value (~3) measurable with our system. In this case, the bandwidth is 1.4 THz (from 0.9 THz to 2.3 THz). In contrast, as shown in Fig. 4a, the imaginary part of the complex refractive index of the CRI mode is close to zero (0.044 at 1.6 THz, which corresponds to an absorption coefficient of 3 cm$^{-1}$, or a 13 % absorption loss with a thickness of 0.63 mm). This enables us to make a crystal thicker to achieve desired bandwidth without sacrificing much of the transmittance of the CRI mode. The loss of the CRI mode shown in Fig. 4b mainly comes from Fresnel losses, which can be significantly reduced by an anti-reflection coating.

To use *n*-InSb as a broadband half-wave plate, the rotation angle (ellipticity) must be constant (zero) within the desired frequency window. This requires the CR frequency to be higher than that frequency window. For example, at 3 T where the CR frequency is about 4.9 THz, the rotation angle is almost constant below 2.5 THz with zero ellipticity as shown in Fig. 3. The small effective mass of electrons (~0.018$m_0$ at 184 K) in InSb makes it easy to satisfy this condition at relatively low magnetic fields. The rotation angle can be set to $\pi/2$ by changing the thickness of the *n*-InSb crystal and fine-tuning magnetic fields.

Finally, we will discuss how the performance of the *n*-InSb polarization optics depends on the material parameters such as the electron density and scattering time. The cold magneto-plasma model, which describes our experimental data very well, provides a guideline for tailoring desired polarization characteristics. For further broadband operation as a circular polarizer, there is a trade-off between the bandwidth and the insertion loss. To achieve ultra-broad bandwidth (>> 1 THz), we need to shorten the scattering time, increase the electron density, or thicken the crystal in an extreme way. However, these lead to a significant increase in the transmission loss of the CRI mode due to the residual absorption from the high-frequency tail of the magneto-plasma resonance at ~ 0.1 THz (Fig. 4a). For less broadband operation with ~1 THz bandwidth, it is easy to achieve very low loss (less than 10 % absorption loss) circular polarizer and the center frequency is tunable with the magnetic field. The working frequency range demonstrated here is from 0.3 to 2.5 THz, which is only limited by the THz detection bandwidth, and is already ~1.5 times broader than that of the previously reported achromatic quarter-wave plate [15]. The *n*-InSb circular polarizer should work in the higher frequency region except for the Reststrahlen band (5.5-5.9 THz). Both the dispersion around the cyclotron resonance and the absorption loss increase with temperature (*T*) due to an exponential increase in the number of thermally excited carriers [20]. As illustrated in the theoretical plot in Fig. 5, at elevated temperatures one can achieve similar performance to that in cooled samples by increasing the magnetic field and decreasing the sample thickness. Figure 5 is plotted for *T* = 250 K, the magnetic field *B* = 5 T and a sample of 0.3 mm thickness. The density of thermally excited carriers is calculated to be 3x10$^{15}$ cm$^{-3}$, which leads to the absorption coefficient around 15 cm$^{-1}$ over the plotted frequency range. The maximum input THz field would be ~50 kV/cm since above this field strength impact ionization is expected to produce high carrier densities (on the order of 10$^{16}$ cm$^{-3}$) [32]. Finally, although the *n*-InSb polarization optics work in a "static" mode with a permanent magnet, a "dynamic" mode is possible with a repetitive pulsed electromagnet [33] to make a fast polarization modulator. Recent advancements in the table-top pulsed electromagnets (peak magnetic fields of around 10 T) will enable us to realize such devices.

## 6. Conclusion

We observed a giant Faraday effect in *n*-InSb using polarization-resolved THz time-domain spectroscopy. Polarization rotation angles and ellipticities as large as $\pi/2$ and 1, respectively were obtained over a wide range of frequencies and were tunable with the external magnetic field, temperature (electron density), and crystal thickness. The results show its promising ability for constructing broadband and tunable THz polarization optics, such as a circular polarizer, half-wave plate, and polarization modulators.

## Acknowledgements

This work was supported by the National Science Foundation through Grants DMR-1006663 and OISE-0968405.